Donald W. Lincoln
Fermi National Accelerator Laboratory

**A brief guide to science communication**

Like many people, I was once a child. And, like many Physics Today readers, I was really interested in all things scientific, from dinosaurs to space travel. But there was a problem. The world in which I grew up was woefully devoid of scientists as role models and sources of information. My parents never went to college; indeed, my father never finished high school. My high school guidance counselor had no clue what a physicist did and how to begin a career in that field.

Luckily, a young science enthusiast in the 1970s had access to the writings of people like Isaac Asimov, Carl Sagan, and George Gamow, who took care to share the world of science with public audiences. And I was a voracious reader, which allowed me to learn from these legendary communicators. Without them, I would likely not have become a physicist. I owed them and others a great debt.

After earning a PhD, I was determined to settle my intellectual tab by paying it forward, hoping to help some other young person living in an academically impoverished environment to view the world with scientific eyes. I began visiting schools and giving tours of my laboratory, and I believe that I sparked interest in a few youngsters. A handful went on to receive PhDs in science. It was quite gratifying.

As the years rolled on, my interest in science outreach broadened. I saw too many examples of public policy that ran afoul of established science to limit my interest in science outreach only to young people. Over the past few decades, I have spent an increasing fraction of my time doing public engagement with other sets of audiences, and I have tried to persuade other scientists to join me in my engagement efforts (see my recent *Physics Today* piece "A defense of science communication" and reference 1).

While I have not been universally successful in convincing my peers, I have encountered a few who also want to share the fascinating principles that govern the behavior of matter and energy in the universe, as well as their own personal journeys into the world of professional science. Some of them have asked me for advice on how to communicate science effectively. This article outlines some of what I've learned over the years.

**Education versus outreach**

So you're interested in doing science communication. The first thing to do is understand exactly what you mean. I find that many physicists conflate the terms "education" and "outreach," but really, they are different things. Education presupposes that the recipients—that is, the students—desire the knowledge the teacher is giving them, whether for the sake of learning, applying the ideas to a career, doing well on a test, or something else. Outreach presupposes no such desire. It is a bit more like advertising, which is to say that you are trying to connect with an audience with little to no prior interest. Both education and outreach are themselves subdivided into even smaller audiences, as illustrated in figure 1.

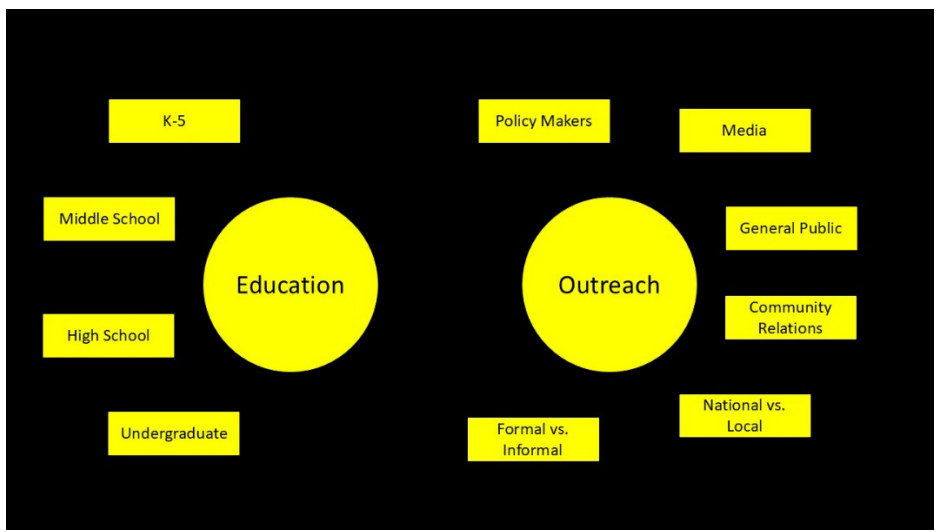

**Figure 1: Education and outreach are distinct efforts and each of these broad categories can be subdivided further into distinct and non-overlapping audiences.**

Given that most scientists are either teachers themselves or at least have ample experience being a student, education is often the easier of the two for them. In education, the audience is expected to put in effort, and accordingly, the task of transferring information requires less work for the presenter.

On the other hand, in outreach, you must grab the audience member's attention and hold it. If you don't, they will flip the page, change the channel, or move on to the next YouTube video.

Of course, the distinction between education and outreach is not entirely black and white. After all, it is difficult to grab the attention of someone who has no interest at all in what you have to say.

Note that you're not ever going to interest all people, so don't try. And if it is your goal to do outreach with people who don't have a preexisting interest in science, realize that that will require considerably more effort and different techniques than you would use for science enthusiasts. For those new to outreach, I recommend beginning with audiences that might be called "sci curious."

The message you want to convey is very important. Many scientists who do outreach have in the back of their head a young version of themselves, and they want to try to nurture a lifelong interest in science in that individual. This sort of education and outreach looks to the distant future. Others have more of an interest in the now and are worried about the vocal and influential antiscience voices that one sees both in society at large and, more worryingly, the corridors of power. The message you want to convey will influence the manner in which you tell it.

The audience you are trying to reach will also influence how you give the message. If you were giving a lecture in the Piazza Navona in Rome and you wanted to communicate effectively, you'd speak in Italian. You must make similar considerations for any audience. If you are speaking with adolescent students, you cannot assume that they know the language of even an introductory physics class. And even many older audience members never learned (or don't remember) Newton's laws. Of course, if your audience is a roomful of retired engineers, you can use the overlapping vocabulary you have with them and don't need to do as much work to make your material seem relevant to them.

Knowing your audience is more than knowing their language. It's knowing their values. You would take a very different approach with a group of high school physics teachers than you would with a community group very focused on social change or a group of retired veterans who might be more interested in funding Veterans Affairs hospitals than scientific research.

Furthermore, the better you know your audience, the better you can tailor both your message and your approach. Figure 2 gives a sense of the diversity of audiences you might encounter if you were interested in lobbying policymakers. Each group has its own language, values, and arenas of interest.

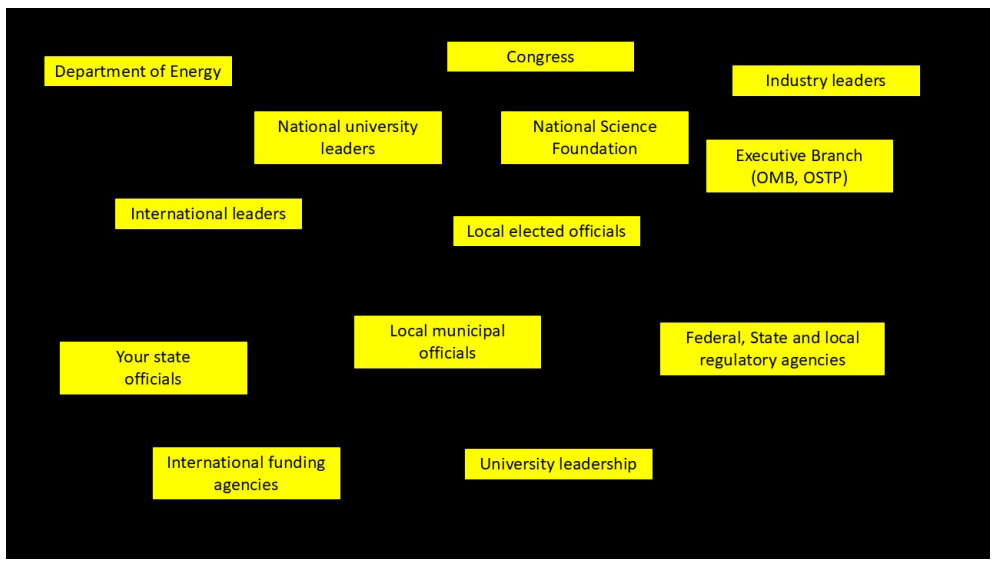

**Figure 2: For a person interested in informing or guiding public policy, there are many smaller audiences, each with their own concerns and sensitivities. If you wish to speak to power, the better you understand the people to whom you're speaking, the more effective you will be.**

I cannot give advice for your specific situation, but here is a checklist of some of the questions you need to ask yourself before undertaking an outreach or education effort.

1. What are you trying to do?
2. What audience do you need to engage with to accomplish your outcome?
3. What language (broadly defined) should you use?
4. Are there cultural sensitivities you should consider? (For example, a religious audience will require a different approach from an atheist one.)
5. Are you speaking to inspire? To inform? To persuade? To call to action?
6. What is the methodology that you wish to employ?
7. How will you know if you are a success?

Answering those last two questions requires knowing what methodologies are effective and what considerations you might encounter for each one.

**Methodologies**

Many readers of *Physics Today* will have many years of experience with education from their times as students and perhaps also as teachers. Furthermore, the American Physical Society has a great deal of resources for educators and offers the ability to join its Forum on Education.[2]. Because of that, I will concentrate my remarks more on the field of public outreach, often called engagement.

There are many methodologies one can employ, including giving public lectures in small or large venues, using traditional media, and attempting to harness the vast potential of the internet. Figure 3 gives a sense of some of the various ways in which one can do outreach. While there is some overlap in the way each approach works, each technique has its own idiosyncrasies. Before you pursue some sort of outreach effort, I suggest that you talk to someone who is successfully doing outreach using a technique that's the same as or similar to the one you envision using. While their efforts may differ somewhat from the one you envision, they will be able to offer advice that applies to you. There is no need to reinvent the wheel.

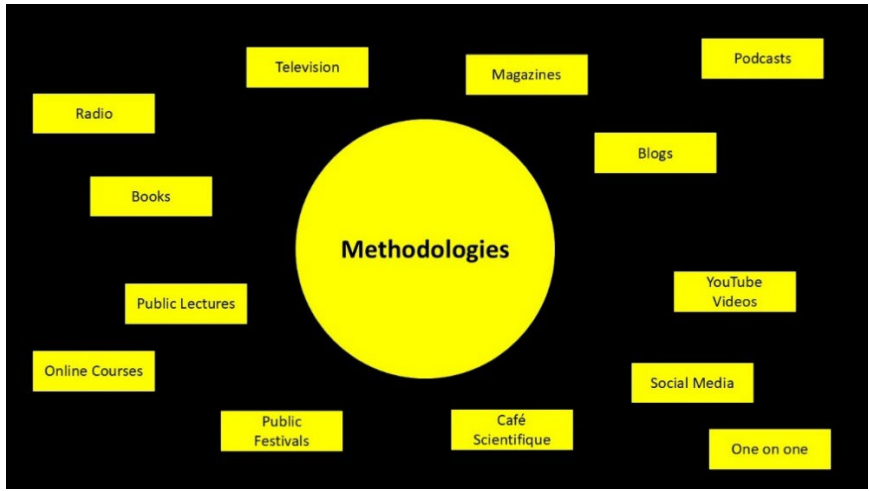

**Figure 3: If you wish to do outreach, first consider how you wish to go about connecting with your desired audience. There are many approaches, and each one of them involves different techniques**

Public talks

Probably the simplest way to do outreach is to give a public talk. This could be done as part of an event (like a conference) or another established effort (like a monthly meeting of a civic or social group). The important thing here is to speak with your host to better understand your audience. Sometimes, venues will host gatherings that combine drinking beverages, socializing, and learning, such as Nerd Nites and Cafes Scientifiques. You can also nominate yourself online to speak at a TED event or contact a TEDx organizer about speaking at their event. Those talks are often posted online as well. (For more about these events and other resources, see the box.)

If you are giving a talk at a recurring social event, it is probably best to attend one or two before you speak, to get the vibe. If you want to start a recurring event of your own, that's a heavier lift. Finding a steady supply of interesting speakers can grow wearisome, and it can take some time to grow your audience. Be ready to be in it for the long haul, and know that organizing is easier if you have a group to share the load.

Lobbying Congress or other policymaking bodies

Lobbying organizations like Congress and other policymaking bodies can be very influential; after all, these are institutions that have the resources and power to effect real change. But how do you convince them to execute some action that you want?

Now, you it could be that you happen to be both charming and persuasive. But charm is not always enough.  What convinces most members of elected bodies are the elections that hire them, and winning elections mean getting votes. Furthermore, it is rare that convincing one person is sufficient to change public policy. Thus, it's most effective to try to influence many policymakers with many voices. That means joining a large group effort, such as the American Physical Society's annual Congressional Visits Day. If you prefer to get your own group going, then perhaps speak to people who have participated in the APS visits or similar efforts to get some pointers.

Art

Art and science are often thought to be diametrically opposed disciplines, one concerned with aesthetics and perceptions and the other revolving around facts and numbers. There are some people who combine the two in what are called STEAM (science, technology, engineering, art, and math) events. Personally, I am not entirely sold that this connection is one of great value to science, although it is certainly true that aesthetic and engaging ways to present data can help persuade skeptical audiences.

I believe, however, that there is definitely a manner in which a combination of art and science can work in physics engagement, and that is to connect with artistic people who have been put off from science at some point in their life. By collaborating with artists of some sort and carefully translating scientific topics into an artistic form, you can teach a little science at art events. It is important to leave the nuts-and-bolts numbers behind and talk about big ideas, but many art–science efforts have had some success. For example, at Yale University, physics professor Sarah Demers has collaborated with dance professor Emily Coates[3] (both are pictured in figure 4), resulting not only in a Yale class on the physics of dance but also public events and even a book.[4]

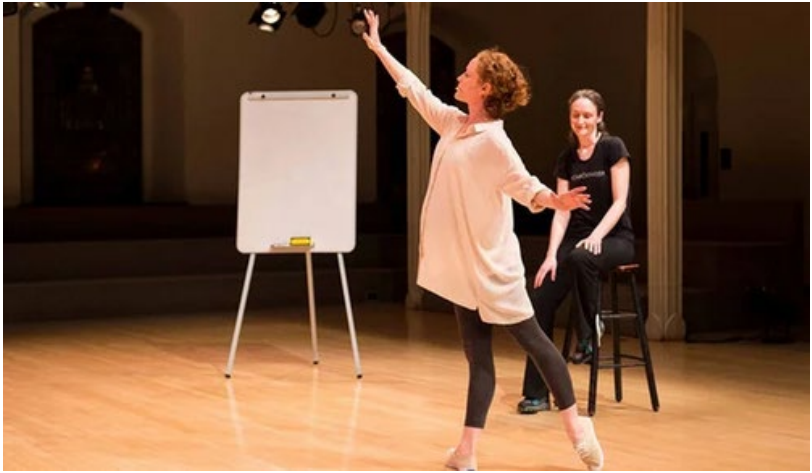

**Figure 4: Dancer Emily Coates (left) and physicist Sarah Demers (right) collaborated to share the physics of dance to a diverse set of audiences, including their "Incarnations" dance presentation, which was featured in the New York Times art section.**

My own laboratory has a guest artist and composer program,[5] and there are similar programs at CERN.[6] I have myself spoken at art events and museums. It can be fun, although it is a different experience from most outreach.

Books

Books are a more conventional way to connect with large audiences, although traditional publishing is becoming increasingly more difficult as the field shrinks. The keys to writing a book are being able to write clearly and engagingly and having something interesting and innovative to say. A popular book is quite different from a textbook, and the competition is fierce. Brian Greene's book *The Elegant Universe: Superstrings, Hidden Dimensions, and the Quest for the Ultimate Theory (*1999) was successful because it was written in the early days of the public awareness of M-theory. Nowadays, a similar book would have a harder time of it. And if you wanted to write about dark matter, you would have to say something new about what has become well-trodden ground.

If you wish to write a popular-science book, you begin with a book proposal,[8] a writing sample (say, a chapter), and a specialized CV that highlights both your expertise and why you're the right person to write the book. You send that to an agent or acquisitions editor. If they accept your proposal, then you write your book.

Publishing a book can be accomplished through self-publishing (I'm not a fan), a university press, or a popular press. A university press is easiest for most physicists. Those presses tend to be looking for more niche books and will accept more-modest sales numbers. Even better, academic credentials will make you more attractive to university presses as an author. For them, you send your proposal package directly to the university press's acquisitions editor.

In contrast, publishing with a popular press, which is to say a traditional publisher not affiliated with a university, usually requires that you have an agent, which can be quite difficult to get. The competition is fierce, and popular presses typically want to see books that will sell a lot of copies.

Except for a lucky few, book publishing is not particularly lucrative. But having a popular-science book published is often a way to demonstrate that you are a serious writer, which can open up other opportunities.

Blogging, social media, online videos

In today's world, many of the most influential popular science voices have a significant online presence. The good thing is that there is a very small initial investment to generating this sort of content. The bad thing is that there is a very small initial investment to generating this sort of content. Low initial investment means that there are countless people out there, all wanting to be heard. Rising above the cacophony is incredibly challenging.

For those who rise to the top, however, the returns can be significant. Given how online search engines work, people with a significant online presence are often found by journalists when they need someone to supply a comment for an article on which they are writing or (very occasionally) when they need someone to appear either on radio or television.

There are several keys to success online. The first, of course, is creating high-quality content: You want to provide a product that is entertaining and insightful. The second is releasing content reliably and regularly. Depending on the outlet, your readers or viewers may expect to see content weekly, daily, or even several times a day.

There are many sorts of ways to have an online presence, including blogs, social media, online videos, and podcasts. But you have to be patient with these methods: Don't expect to immediately become a viral phenom. It's a long and difficult grind. Initially, you will wonder if anyone is paying attention, a sentiment illustrated in figure 5. One my favorite sayings about this sort of thing is that it takes 10 years of soul-crushing and persistent effort to become an overnight sensation.

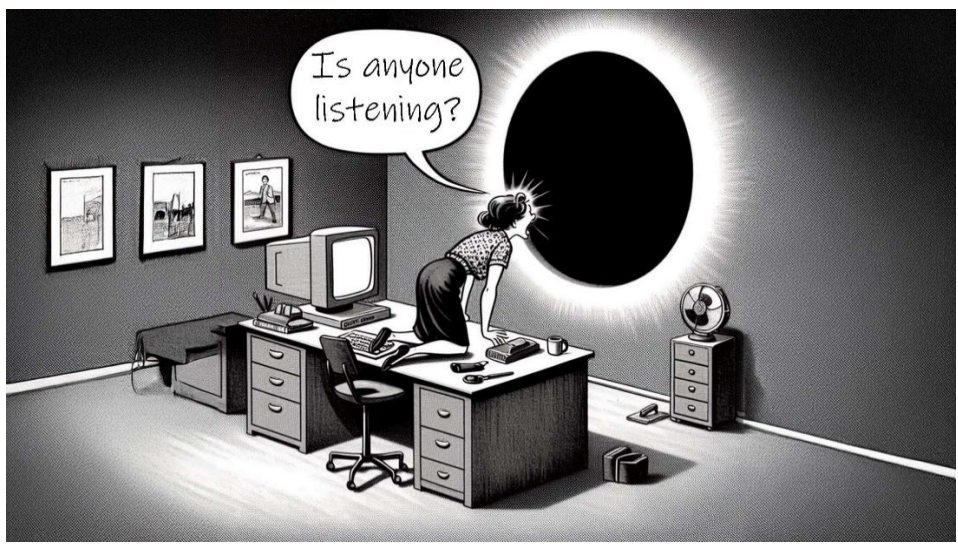

**Figure 5: The initial phase of attempting to do online science outreach can be very depressing.**

Success online benefits greatly from connections. A guest appearance or mention on a popular website can do wonders for your viewership numbers. When I began my entrée into the world of social media well over a decade ago, it took years to grow my Facebook following to 500 people. Then one mention on one successful site added

1000 followers overnight. My current following consists of nearly 30 000—a small number compared with professional communicators but a reasonable one for a person who continues to stay connected to the research world.

Each online platform is quite different. Podcasts and YouTube can require a larger monetary investment than many others do, as they require recording equipment and editing software. Producing a high-quality episode is considerable work and, if your approach involves interviewing others, you'll have the constant grind of finding guests.

For social media, the entrée is easier, but you need to know the personality of your platform. Not all messages are well matched to all platforms. Facebook is great for sharing material, and the demographic skews older than the youth-friendly Instagram, which is more of a visual and image-centric platform.

In addition, you should consider recent evolutions in the world of social media. Twitter (now X) was once far more influential than it is now. The popularity of Snapchat has waned, especially in North America. TikTok has been on the rise. Expect to constantly reinvent yourself.

Furthermore, you should take regional preferences into consideration. For example, Telegram is popular in parts of Europe, Asia, and Africa, while WhatsApp is popular in those locations and parts of South America. In contrast, neither of them is dominant in North America. Depending on the audience you want to reach, you should pick your platform carefully.

**Big picture**

Attempting to communicate with the public can be a daunting prospect. Many people are indifferent to science. Some may have learned "the mitochondria is the powerhouse of the cell" in school but not understand what science really is: a way of figuring things out. Others are frankly hostile toward the scientific enterprise. The world is rife with mis- and disinformation. If fighting that deluge is your goal, it can feel like a thankless, never-ending game of Whac-A-Mole.

And yet science outreach can be greatly rewarding. You can open new vistas to young people who will one day be scientists. You can shape public opinion and nudge science policy in the direction of research and reason. On a practical level, you can positively benefit public funding of science and, more self-centeredly, persuade funding agencies to support the work that interests you.

If you're inclined to do science outreach, I hope you start. If you'd prefer to let others take on that burden, that's OK too, but you should be supportive of—indeed, grateful for—their work. After all, effective communicators are making society more open to hearing about your research or public officials more likely to support you when you ask for resources.

We live in a connected world, with a sometimes-deafening hubbub of voices. We should work together to ensure that the voice of science is heard.